\documentstyle[12pt]{article} 
\begin{document}

\title{Some Considerations on Lattice Gauge Fixing}
\author{Massimo Testa \\
Dip. di Fisica, Universit\'{a} di Roma
``La Sapienza``\\ and\\ INFN Sezione di Roma I\\
Piazzale Aldo Moro 2\\I-00185 Roma\\
 ITALY\\ massimo.testa@roma1.infn.it}
\maketitle
\begin{abstract}
Some problems related to Gribov copies in lattice gauge-fixing  and their possible solution are discussed.
\end{abstract}

\section{Gribov and Gauge-Fixing Problem} \label{00}

The Faddeev-Popov\cite{f-p} quantization gives a meaning to the 
formal (euclidean)
expectation value of a gauge invariant observable operator:
\begin{equation}
\left\langle O \right\rangle ={{\int {\delta A\exp [-S(A)]}\;O(A)} 
\over {\int {\delta A\exp [-S(A)]}}}	
	\label{uno}
\end{equation}¥
The Faddeev-Popov method requires the choice of a gauge fixing condition:
\begin{equation}
f(A)=0	
	\label{due}
\end{equation}¥
in terms of which we define $\Delta (A)$ as:
\begin{equation}
\Delta (A)\cdot \int {D\Omega \delta [f(A^\Omega )]}=1	
	\label{tre}
\end{equation}¥
In eq.(\ref{tre}), $D\Omega$ denotes the invariant measure on the 
gauge group $G$.
It is easy to show that $\Delta (A)$ is gauge invariant:
\begin{equation}
\Delta (A^\Omega )=\Delta (A)	
	\label{cinque}
\end{equation}¥
We then get the following gauge-fixed expression 
for $\left\langle O \right\rangle$:
\begin{equation}
\left\langle O \right\rangle ={{\int {\delta A\exp [-S(A)]\Delta 
(A)\delta [f(A)]O(A)}} \over {\int {\delta A\exp [-S(A)]\Delta 
(A)\delta [f(A)]}}}	
	\label{sei}
\end{equation}
Let us consider, at the moment, the academic situation in which Gribov 
copies\cite{grib} are absent. Then, if $A$ satisfies $f(A)=0$, we 
get an explicit expression for $\Delta(A)$:
\begin{equation}
\Delta (A)=\left. {\det {{\delta f(A^\Omega )} \over {\delta \Omega 
}}} \right|_{\Omega =1}	
	\label{quattro}
\end{equation}¥
Choosing, e.g., $f(A)=\partial _\mu A^\mu $ as the gauge fixing 
condition, we have:
\begin{equation}
\left\langle O \right\rangle ={{\int {\delta A\exp [-S(A)]\det [\partial D(A)]
\delta [\partial _\mu A^\mu]O(A)}} \over {\int {\delta A\exp [-S(A)]
\det [\partial D(A)] \delta [\partial _\mu A^\mu]}}}	
	\label{seibis}
\end{equation}
More generally we can eliminate the $\delta $-function from the 
functional integrand as:
\begin{equation}
\left\langle O \right\rangle ={{\int {\delta Ae^{[-S(A)-{1 \over {2\alpha }}\int 
{(\partial A)^2}]}\det [\partial D(A)]} O(A)} \over
{\int {\delta Ae^{[-S(A)-{1 \over {2\alpha }}\int 
{(\partial A)^2}]}\det [\partial D(A)]}}}
	\label{sette}
\end{equation}
Eq.(\ref{sette}) can be linearized through the introduction of the Lagrange 
multipliers $\lambda(x)$:
\begin{equation}
\left\langle O \right\rangle={{\int {\delta A\delta
\lambda e^{[-S(A)+i\int {\lambda \partial 
A}-{\alpha  \over 2}\int {\lambda ^2}]}\det [\partial D(A)]})O(A)} 
\over {\int {\delta A\delta \lambda e^{[-S(A)+i\int {\lambda \partial 
A}-{\alpha  \over 2}\int {\lambda ^2}]}\det [\partial D(A)]})}}
	\label{settebis}
\end{equation}
It is now possible to rewrite eqs.(\ref{sette}),(\ref{settebis})
introducing ghost and antighost fields, $c(x)$ and $\bar c(x)$:
\begin{eqnarray}
\left\langle O \right\rangle =
{{\int {\delta A\delta c\delta \bar ce^{[-S(A)-{1 \over {2\alpha 
}}\int {(\partial A)^2}-\int {\partial \bar cD(A)c}]}}O(A)} \over
{\int {\delta A\delta c\delta \bar ce^{[-S(A)-{1 \over {2\alpha 
}}\int {(\partial A)^2}-\int {\partial \bar cD(A)c}]}}}}=
\label{ottobis} \\
= {{{\int {\delta A\delta \lambda \delta c\delta \bar ce^{[-S(A)+i\int {\lambda \partial 
A}-{\alpha  \over 2}\int {\lambda ^2}-\int {\partial \bar cD(A)c}]}}}O(A)} \over
{\int {\delta A \delta \lambda \delta c\delta \bar ce^{[-S(A)+i\int {\lambda \partial 
A}-{\alpha  \over 2}\int {\lambda ^2}]-\int {\partial \bar 
cD(A)c}]}}}}
\label{otto}
\end{eqnarray}¥
In the formulation given in eq.(\ref{otto}), the theory admits a
nilpotent ($\delta ^2=0$) BRST symmetry\cite{brst}: 
\begin{eqnarray}
\delta A_\mu =D_\mu (A)c \nonumber \\
\delta \bar c=i\lambda  \nonumber \\
\delta c={1 \over 2}cc \nonumber \\
\delta \lambda =0 \label{nove}
\end{eqnarray}¥
BRST invariance follows from the fact that eq.(\ref{otto}) can be 
rewritten, through eqs.(\ref{nove}), as:
\begin{equation}
\left\langle O \right\rangle ={1 \over {Z'}}\int {D\mu 
e^{-S(U)}e^{-{\alpha  \over 2}\int {\lambda ^2}}e^{\delta \int 
{f\,\bar c}}O(U)}	
	\label{dieci}
\end{equation}
where:
\begin{equation}
Z'=\int {D\mu e^{-S(U)}e^{-{\alpha  \over 2}\int {\lambda 
^2}}e^{\delta \int {f\,\bar c}}}	
	\label{undici}
\end{equation}
and:
\begin{equation}
D\mu \equiv DAd\lambda d\bar cdc	
	\label{dodici}
\end{equation}
is the BRST-invariant measure.

Gribov copies correspond to multiple $\Omega$ solutions of the equation
\begin{equation}
f(A^{\Omega})=0	
	\label{tredici}
\end{equation}
for a given a gauge field configuration $A_{\mu}^{a}(x)$.
Labeling the different solutions of eq.(\ref{tredici}) by $\Omega _i$, 
we get from eq.(\ref{tre}):
\begin{equation}
\Delta (A)^{-1}=\sum\limits_i {{1 \over {\left| {\det {{\delta 
f(A^\Omega )} \over {\delta \Omega }}} \right|_{\Omega _i}}}}	
	\label{quattordici}
\end{equation}¥
Although correct, the use of eq.(\ref{quattordici}) is very 
inconvenient: in this way the ghost formulation is lost and, with it,
the related BRST invariance.

An alternative procedure which maintains BRST 
symmetry also in presence of Gribov copies, has been suggested long
ago\cite{sol}. It consists in the observation that the quantity:
\begin{equation}
n(A)\equiv \int {D\Omega \det \left[ {{{\delta f(A^\Omega )} \over 
{\delta \Omega }}} \right]\delta [f(A^\Omega )]}	
	\label{quindici}
\end{equation}¥
is a sum of $\pm 1$, which counts the intersections
(with sign) of the gauge orbit with the gauge fixing surface. As a consequence
of an index theorem, it turns out that $n(A)$ is independent of $A_\mu $. 
If we could show, in addition, that $n(A) \ne 0$, the Faddeev-Popov 
formulation and the BRST symmetry would follow at once. In particular,
if the gauge condition $f(A)=\partial _\mu A^\mu $ were chosen,
we would get precisely eqs.(\ref{ottobis}),(\ref{otto}) even in 
presence of Gribov copies.

Lattice regularization\cite{wil} offers a unique opportunity to study this 
problem, since, due to the compactness of the lattice gauge fields,
$U_\mu $, both the gauge fixed and non gauge fixed versions of the path
integral are meaningful.
However, as explained in section \ref{01}, precisely because of compactness,
we can show\cite{neub1} that $n(A)=0$.

\section{Neuberger problem ($n(A)=0$)} \label{01}

We start with the lattice regularized expectation value of a gauge
invariant operator $O(U)$:
\begin{equation}
\left\langle O \right\rangle ={1 \over Z}\int {DUe^{-S(U)}O(U)}	
	\label{sedici}
\end{equation}¥
where:
\begin{equation}
Z=\int {DUe^{-S(U)}}	
	\label{diciassette}
\end{equation}¥
It is easy to introduce a gauge-fixing $f(A)$ along the same lines 
discussed in section \ref{01}:
\begin{equation}
\left\langle O \right\rangle ={1 \over {Z'}}\int {D\mu 
e^{-S(U)}e^{-{\alpha  \over 2}\int {\lambda ^2}}e^{\delta \int 
{f\,\bar c}}O(U)}	
	\label{diciotto}
\end{equation}¥
where:
\begin{equation}
Z'=\int {D\mu e^{-S(U)}e^{-{\alpha  \over 2}\int {\lambda 
^2}}e^{\delta \int {f\,\bar c}}}	
	\label{diciannove}
\end{equation}¥
and:
\begin{equation}
D\mu \equiv DUd\lambda d\bar cdc	
	\label{venti}
\end{equation}¥
In eqs.(\ref{diciotto}),(\ref{diciannove}), $\delta$ denotes a
nilpotent ($\delta ^2=0$) Lattice BRST transformation defined by:
\begin{eqnarray}
\delta U_\mu =c(x)U_\mu (x)-U_\mu (x)c(x+a\hat \mu ) \label{na} \\
\delta \bar c=i\lambda \\
\delta c={1 \over 2}cc \\
\delta \lambda =0
\end{eqnarray}¥
Following \cite{neub1}, we can now define:
\begin{equation}
F_O(t)\equiv \int {D\mu e^{-S(U)}e^{-{\alpha  \over 2}\int {\lambda 
^2}}e^{t\delta \int {f\,\bar c}}O(U)}	
	\label{ventidue}
\end{equation}¥
so that, due to nilpotency:
\begin{eqnarray}
{{dF_O(t)} \over {dt}}\equiv \int {D\mu [\delta \int 
{f\,\bar c]}e^{-S(U)}e^{-{\alpha  \over 2}\int {\lambda 
^2}}e^{t\delta \int {f\,\bar c}}O(U)}=\\
=\int {D\mu \delta [\int {f\,\bar c\;}e^{-S(U)}e^{-{\alpha  \over 
2}\int {\lambda ^2}}e^{t\delta \int {f\,\bar c}}O(U)}]=0	
	\label{ventitre}
\end{eqnarray}
On the other hand we also have:
\begin{equation}
F_O(0)=\int {DUd\lambda d\bar cdc\;e^{-S(U)}e^{-{\alpha  \over 
2}\int {\lambda ^2}}O(U)}=0	
	\label{ventiquattro}
\end{equation}¥
as a consequence of Berezin integration rules, since the integrand of 
eq.(\ref{ventiquattro}) does not contain ghost, nor antighost fields.

We get therefore:
\begin{equation}
F_O(1)=\int {D\mu e^{-S(U)}e^{-{\alpha  \over 2}\int {\lambda 
^2}}e^{\delta \int {f\,\bar c}}O(U)=}0	
	\label{venticinque}
\end{equation}
and the expectation value of any observable assumes the form
$\left\langle O \right\rangle ={0 \over 0}$.

As discussed in section \ref{02}, this situation is the consequence of a
cancellation among Lattice Gribov copies.

\section{Toy Abelian Model} \label{02}
In this section we will consider a zero dimensional prototype of 
abelian BRST
symmetry with compact variables\cite{toy} which will clarify the 
nature of the problem and a possible way out.
The model consists of one "link" variable $U$,
which we choose to parametrize through its phase, as:
\begin{equation}
U=e^{iaA}	
	\label{ventisei}
\end{equation}¥
where:
\begin{equation}
-{\pi \over a} \le A \le {\pi \over a}
	\label{ventisette}
\end{equation}¥
$a$ is a parameter, reminiscent of
the lattice spacing in more realistic situations, whose limit
$a \rightarrow 0$ will be used to connect the periodic, compact case to the
non compact one.

We define:
\begin{equation}
N\equiv \int\limits_{-{\pi  \over a}} ^{{\pi \over  a}} {dA}	
	\label{ventotto}
\end{equation}¥

The Ògauge-fixedÓ version of the ``functional'' integral in 
eq.(\ref{ventotto}), is:
\begin{equation}
N'=\int\limits_{-{\pi  \over a}}^{{\pi  \over a}} 
{dA\int\limits_{-\infty }^{+\infty } {d\lambda \int {d\bar 
cdce^{-{\alpha  \over 2}\lambda ^2}e^{\delta [\bar cf(A)]}}}}	
	\label{ventinove}
\end{equation}¥
where $\delta $ denotes the (nilpotent $\delta ^2=0$) BRST-like transformation:
\begin{eqnarray}
\delta A=c \nonumber \\
\delta c=0 \nonumber \\
\delta \bar c=i \lambda \nonumber \\
\delta \lambda =0	
	\label{trenta}
\end{eqnarray}¥
Going through the same steps as in section \ref{01}, we conclude that
$N'$ suffers from the Neuberger disease:
\begin{equation}
N'=0	
	\label{trentuno}
\end{equation}¥
This can also be checked through an explicit calculation:
\begin{eqnarray}
N'=\int\limits_{-{\pi  \over a}}^{{\pi  \over a}} 
{dA\int\limits_{-\infty }^{+\infty } {d\lambda \int {d\bar 
cdce^{-{\alpha  \over 2}\lambda ^2}e^{i\lambda f(A)}e^{-\bar 
cf'(A)c}}}}= \nonumber \\
=\int\limits_{-{\pi  \over a}}^{{\pi  \over a}} 
{dA\int\limits_{-\infty }^{+\infty } {d\lambda \;e^{-{\alpha  \over 
2}\lambda ^2}e^{i\lambda f(A)}f'(A)}}= \nonumber \\
=\sqrt {{{2\pi } \over \alpha }}\int\limits_{-{\pi  \over 
a}}^{{\pi  \over a}} {df(A)e^{-{{f(A)^2} \over {2\alpha }}}}=0	
	\label{trentadue}
\end{eqnarray}¥
for a periodic, non-singular, $f(A)$.
The reason why we need a periodic $f(A)$ is that we want BRST Identities
to be satisfied. This is crucial to show independence on $\alpha $ of
gauge-invariant observables. The prototype of BRST Identities is:
\begin{equation}
\left\langle {\delta \Gamma } \right\rangle =0	
	\label{trentatre}
\end{equation}¥
where $\Gamma$ is any quantity with ghost number $-1$.
If we choose:
\begin{equation}
\Gamma \equiv \bar cF(A,\lambda)	
	\label{trentaquattro}
\end{equation}¥
so that:
\begin{equation}
\delta \Gamma \equiv \delta [\bar cF(A,\lambda)]=i\lambda F(A,\lambda)-\bar 
cF'(A,\lambda)c
	\label{trentacinque}
\end{equation}
where the $'$ denotes the derivative with respect to $A$,
we have:
\begin{eqnarray}
\left\langle {\delta \Gamma } \right\rangle \equiv 
\int\limits_{-{\pi  \over a}}^{{\pi  \over a}} 
{dA\int\limits_{-\infty }^{+\infty } {d\lambda \int {d\bar 
cdce^{-{\alpha  \over 2}\lambda ^2}e^{\delta [\bar cf(A)]}\delta 
\Gamma }}}= \nonumber \\
=\left.\int\limits_{-\infty }^{+\infty } {d\lambda
e^{-{\alpha  \over 2}\lambda ^2}e^{i\lambda f(A)]}
F(A,\lambda)}\right|_{A=-{\pi  \over a}}^{A={\pi  \over a}}=0	
	\label{trentasei}
\end{eqnarray}¥
which can only be satisfied for a periodic (in $A$) Ògauge fixing conditionÓ,
$f(A)$, and $F(A,\lambda)$.
In particular, for $\alpha =0$, eq.(\ref{ventinove}) becomes:
\begin{eqnarray}
N'=\mathop {\lim }\limits_{\alpha \to 
0}\int\limits_{-{\pi  \over a}}^{{\pi  \over a}} 
{dA\int\limits_{-\infty }^{+\infty } {d\lambda \;e^{-{\alpha  \over 
2}\lambda ^2}e^{i\lambda f(A)}f'(A)}}= \nonumber \\
=2\pi \int\limits_{-{\pi  \over a}}^{{\pi  \over a}} 
{dA\;f'(A)\;\delta \left( {f(A)} \right)}=0	
	\label{trentasette}
\end{eqnarray}¥
which displays the Gribov nature of the paradox: a periodic 
$f(A)$ has an even number of zeroes which contribute 
alternatively $\pm 1$ to eq.(\ref{trentasette}) and cancel exactly.

Within this toy abelian model, the solution of the ''Gribov problem''
is simple. It is enough to substitute the gauge fixing 
$\delta$-function with a periodic $\delta$-function\cite{light}:
\begin{equation}
\delta \Rightarrow \delta _P	
	\label{trentotto}
\end{equation}¥
with:
\begin{equation}
\delta _P(x)\equiv \sum\limits_{n=-\infty }^{+\infty } 
{\delta (x-n{{2\pi } \over a})}={a \over {2\pi 
}}\sum\limits_{n=-\infty }^{+\infty } {e^{inax}}\equiv {a \over {2\pi 
}}\sum\limits_{n=-\infty }^{+\infty } {e^{i\lambda _nx}}	
	\label{trentanove}
\end{equation}¥
In eq.(\ref{trentanove}) we put:
\begin{equation}
 \lambda _n\equiv na
	\label{quaranta}
\end{equation}¥

Extending the formulation with the inclusion of a $\lambda_{n}^2$ at the 
exponent, analogous to eq.(\ref{ventinove}), we then have: 
\begin{equation}
N'\Rightarrow N''	
	\label{quarantuno}
\end{equation}¥
where:
\begin{equation}
N''=a\sum\limits_{n=-\infty }^{+\infty } {\int\limits_{-{\pi  \over 
a}}^{{\pi  \over a}} {dAe^{-{\alpha  \over 2}\lambda _n^2}e^{i\lambda 
_nf(A)}f'(A)}}	
	\label{quarantadue}
\end{equation}¥

This formulation admits an obvious BRST invariance under transformations
similar to  eq.(\ref{trenta}), provided we interpret the variation of the 
antighost as:
\begin{equation}
\delta \bar c=i\lambda _n	
	\label{quarantatre}
\end{equation}¥
We have, in analogy with eq.(\ref{ventinove}):
\begin{equation}
N''=a\sum\limits_{n=-\infty }^{+\infty } {\int\limits_{-{\pi  \over 
a}}^{{\pi  \over a}} {dA\int {d\bar cdce^{-{\alpha  \over 2}\lambda 
_n^2}e^{\delta [\bar cf(A)]}}}}	
	\label{quarantaquattro}
\end{equation}¥
Invariance under these modified BRST transformations is enough for 
all purposes related to gauge invariance.

The advantage of having a discretize set of Lagrange 
multipliers $\lambda_{n}$, eq.(\ref{quaranta}), is that we are now free
to chose a non-periodic ''gauge fixing'' condition  $f(A)$ such that:
\begin{equation}
f(A+{{2\pi } \over a})=f(A)+{{2\pi } \over a}	
	\label{quarantacinque}
\end{equation}¥
still respecting BRST Identities, eq.(\ref{trentatre}). In fact, while
the integrand of eq.(\ref{quarantadue}) is still periodic, the condition
stated in eq.(\ref{quarantacinque}) evades the cancellation among Gribov copies
because $f(A)$ has an odd number of zeroes. Another way of stating this 
fact is to recognise that $\exp (it\lambda _nf(A))$
is only periodic for integer $t$'s and Neuberger's argument, which 
requires taking a derivative with respect to $t$, is avoided.
When $a\to 0$ we recover the continuum BRST formulation in analogy to
the way in which we get the Fourier integral from the Fourier series:
\begin{eqnarray}
\mathop {\lim }\limits_{a\to 0}N''=\mathop {\lim 
}\limits_{a\to 0}a\sum\limits_{n=-\infty }^{+\infty } 
{\int\limits_{-{\pi  \over a}}^{{\pi  \over a}} {dAe^{-{\alpha  \over 
2}\lambda _n^2}e^{i\lambda _nf(A)}f'(A)}}= \nonumber \\
=\int\limits_{-{\pi  \over a}}^{{\pi  \over a}} 
{dA\int\limits_{-\infty }^{+\infty } {d\lambda \;}e^{-{\alpha  \over 
2}\lambda ^2}e^{i\lambda f(A)}f'(A)}	
	\label{quarantasei}
\end{eqnarray}¥

\section{$U(1)$ Gauge Theory}  \label{03}

The case of the $U(1)$ Gauge Theory can be immediately treated along 
the lines of the toy model. We parametrize the gauge field
$U_\mu(x)$ as:
\begin{equation}
U_\mu (x)\equiv e^{iaA_\mu (x)}	
	\label{quarantasette}
\end{equation}¥
The BRST variation of $A_\mu (x)$, induced by eq.(\ref{na}) is:
\begin{equation}
\delta A_\mu (x)={{c(x+a\hat \mu )-c(x)} \over a}	
	\label{quarantotto}
\end{equation}¥
and we can chose, for example, a discretization of the Lorentz 
gauge-fixing:
\begin{equation}
f(A)=\sum\limits_\mu  {{{[A_\mu (x+a\hat \mu )-A_\mu (x)]} \over 
a}}	
	\label{quarantanove}
\end{equation}¥
In this case:
\begin{equation}
\int {\delta A\delta c\delta \bar c\sum\limits_{{n(x)}} 
{e^{[-S(A)+ia^4\sum\limits_x {\lambda _n(x)\sum\limits_\mu  {{{[A_\mu 
(x+a\hat \mu )-A_\mu (x)]} \over a}}}-{\alpha  \over 
2}a^4\sum\limits_x {\lambda _n^2(x)-\int {\partial \bar c\partial 
c}}]}}}	
	\label{cinquanta}
\end{equation}¥
where:
\begin{equation}
\lambda _n(x)={{n(x)} \over {a^2}}	
	\label{cinquantuno}
\end{equation}¥

\section{Conclusions}

The Fujikawa-Hirschfeld-Sharpe proposal\cite{sol} seems to be viable, at 
least in the abelian compact case. More work is needed to clarify the 
considerably more difficult case of non abelian compact gauge fields. 

\section*{Acknowledgements}
I want to thank the organizers of the Workshop on ''Lattice Fermions and
Structure of the Vacuum'' and in particular Professor Valya 
Mitrjushkin for the generous hospitality and the wonderful organization.

\end{document}